\documentclass{llncs}

\usepackage{etex}

\usepackage{amssymb,amsmath}
\usepackage{xcolor}
\usepackage{paralist}
\usepackage{hyperref}
\usepackage{galois}
\usepackage{verbatim}
\usepackage{mathrsfs}
\usepackage{graphicx}
\usepackage{caption}
\usepackage{subcaption}

\usepackage{tikz}
\usetikzlibrary{calc}
\usetikzlibrary{shapes}
\usetikzlibrary{shapes.multipart}
\usetikzlibrary{shadows}
\usetikzlibrary{arrows}
\usetikzlibrary{patterns}
\usepackage{pgfplots}

\usepackage[ruled,linesnumbered,vlined]{algorithm2e}
\SetKwFor{Func}{function}{}{}
\SetKw{Continue}{continue}

\usepackage{listings}
\usepackage{pstricks,pst-node,pst-plot,pstricks-add}
\newcommand{\dkcmt}[1]{}
\newcommand{\mbcmt}[1]{}
\newcommand{\pscmt}[1]{}
\newcommand{\sjcmt}[1]{}

\newcommand{\pponly}[1]{{}}
\newcommand{\rronly}[1]{{#1}}

\newcommand{\true}{\mathit{true}}
\newcommand{\false}{\mathit{false}}

\renewcommand{\vec}[1]{{\boldsymbol{#1}}}

\newcommand{\vecv}[2]{\left(\begin{array}{c}{#1}\\{#2}\end{array}\right)}

\newcommand{\templ}{\mathcal{T}}
\newcommand{\basetempl}{\widehat{\mathcal{T}}}
\newcommand{\rowexpr}{\mathit{e}}

\newcommand{\trans}{\mathit{Trans}}
\newcommand{\inv}{\mathit{Inv}}
\newcommand{\kinv}{\mathit{KInv}}
\newcommand{\ainv}{\mathit{AInv}}
\newcommand{\start}{\mathit{Start}}
\newcommand{\err}{\mathit{Err}}

\newcommand{\adom}{\mathscr{A}}
\newcommand{\tdom}{\mathscr{T}}

\newcommand{\sprop}{\mathit{P}}
\newcommand{\strans}{\mathit{T}}
\newcommand{\sinv}{\mathit{I}}
\newcommand{\skinv}{\mathit{KInv}}

\newcommand{\sstart}{\mathit{Start}}
\newcommand{\serr}{\mathit{Err}}

\newcommand{\algorithmName}[0]{$k$I$k$I}
\newcommand{\systemName}[0]{2LS}

\newcommand{\define}[1]{\emph{#1}}

\newcommand{\limplies}[0]{\Rightarrow}

\newcommand{\secondorderexists}[0]{\exists_2}
\newcommand{\st}[0]{\centerdot}

\newcommand{\lte}[0]{\leqslant}

\newcommand{\bigland}[1]{\bigwedge_{#1}}

\newcommand{\result}[1]{{\small \color{gray} #1}}

\begin{document}

\title{
Safety Verification and Refutation by 
$k$-invariants and
$k$-induction\rronly{ (extended version)}%
\thanks{This research was supported by
    the ARTEMIS Joint Undertaking under grant
    agreement number 295311 (\href{http://vetess.eu/}{VeTeSS}),
    the Toyota Motor Corporation and 
    ERC project~280053 (CPROVER).    
}}
\author{
Martin Brain, Saurabh Joshi, Daniel Kroening, Peter Schrammel
}
\institute{University of Oxford, UK}

\maketitle

\begin{abstract}
Most software verification tools can be classified into one of a
number of established families, each of which has their own focus and
strengths.  For example, concrete counterexample generation in
model checking, invariant inference in abstract interpretation and
completeness via annotation for deductive verification.
\
This creates a significant and fundamental usability problem as users may have to
learn and use one technique to find potential problems but then need
an entirely different one to show that they have been fixed.
\
This paper presents a single, unified algorithm
\algorithmName, which strictly generalises
abstract interpretation, bounded model checking and $k$-induction.
\
This not only combines the strengths of these techniques but allows
them to interact and reinforce each other, giving a `single-tool'
approach to verification.
\end{abstract}

\section{Introduction}

The software verification literature contains a wide range of techniques
which can be used to prove or disprove safety properties.  
These include:

\begin{description}
  \item[Bounded Model Checking]{Given sufficient time and resource, BMC
    will give counterexamples for all false properties,
    which are often of significant value for understanding the fault.  
    However only a small proportion of true properties can
    be proven by BMC.}

 \item[$k$-Induction]{Generalising Hoare logic's ideas of loop
   invariants, $k$-induction can prove true properties, and, in some
   cases provide counterexamples to false ones.
   However it requires inductive invariants, which can be expensive (in
   terms of user time, expertise and maintenance).}

 \item[Abstract Interpretation]{The use of over-approximations makes
   it easy to compute invariants which allow many true
   propositions to be proven.  However false properties and
   true-but-not-provable properties may be indistinguishable.
   Tools may have limited support for a more complete analysis.}
\end{description}

The range and variety of tools and techniques available is
a sign of a healthy and vibrant research community but
presents challenges for non-expert users.
\
\emph{The choice of which tools to use and where to expend effort depends on
  whether the properties are true or not -- which is exactly what they
  want to find out.}

To build a robust and usable software verification system it is
necessary to combine a variety of techniques.  One option would be to
run a series of independent tools, in parallel (as a portfolio, for
example) or in some sequential order.  However this limits the
information that can be exchanged between the algorithms --
what is needed is a genuine compound rather than a simple mixture.
Another option would be to use monolithic algorithms such as
CEGAR~\cite{CGJ+00}, IMPACT~\cite{Mcm06} or IC3/PDR~\cite{BM07,HB12}
which combine some of the \emph{ideas} of simpler systems. These are
difficult to implement well as their components interact in complex
and subtle ways.  Also they require advanced solver features such as
interpolant generation that are not widely available for all theories
(bit-vectors, arrays, floating-point, etc.).  In this paper, we argue for a
compound with simple components and well-understood interaction.

This paper draws together a range of well-known techniques and
combines them in a novel way so that they strengthen and reinforce each other.
$k$-induction \cite{SSS00} uses
syntactically restricted or simple invariants (such as those generated
by abstract interpretation) to prove safety.  Bounded
model checking~\cite{BCCZ99} allows us to test $k$-induction failures to see
if they are real counter-examples or, if not, to build up a set of
assumptions about system behaviour.  Template-based abstract
interpretation is used for invariant inference~\cite{SSM05,RSY04,GSV08}
with unrolling producing progressively stronger invariants.
Using a solver and templates to generate
invariants allows the assumptions to be used without the need for
backwards propagators and `closes the loop' allowing the techniques to
strengthen each other.
Specifically, the paper makes the following contributions:

\begin{compactenum}
\item{A new, unified, simple and elegant algorithm, \algorithmName,
  for integrated invariant inference and counterexample generation is
  presented in Section~\ref{section:algorithm-concepts}.  Incremental
  bounded model checking, $k$-induction and classical
  over-approximating abstract interpretation are shown to be
  restrictions of \algorithmName.}

\item{The techniques required to efficiently implement
  \algorithmName\ are given in Section~\ref{section:algorithm-details}
  and an implementation, {\systemName}, is described in Section
  \ref{section:implementation}.}

\item{A series of experiments are given in 
  Section~\ref{section:experiments}. We show that {\algorithmName}
  \emph{verified more programs and is faster} than a portfolio
  approach using incremental BMC, $k$-induction and abstract
  interpretation, showing genuine synergy between components. }

\end{compactenum}

\section{Algorithm Concepts}
\label{section:algorithm-concepts}

This section reviews the key concepts behind \algorithmName.  Basic
familiarity with transition systems and first and second order logic
will be assumed.
As we intend to use \algorithmName\ to verify software using bit-vectors,
we will focus on finite state systems.

\subsection{Program Verification as Second Order Logic}\label{sec:pv2ls}

To ease formalisation we view programs as symbolic transition
systems.  The state of a program is described by a logical
interpretation with logical variables corresponding to each program
variable, including the program counter.  Formulae can be used to
describe sets of states -- the states in the set are the models of
the formulae.
Given $\vec{x}$, a vector of variables, $\start(\vec{x})$ is the
predicate describing the start states.
A \define{transition relation}, $\trans(\vec{x},\vec{x'})$ is formula
describing a relation between pairs of such interpretations which
describes the (potentially non-deterministic) progression relations
between states.
From these we
can derive the set of reachable states as the least fixed-point of the
transition relation starting from the states described by $\start(\vec{x})$.
Although this set is easily defined, computing a predicate that
describes it (from $\start$ and $\trans$) is often difficult and we will
focus on the case when it is not practical.  Instead 
\define{inductive invariant} are used;  $\inv$ is an inductive
invariant if it has the following property:
\begin{equation}
\forall \vec{x}_0, \vec{x}_1 \st (\inv(\vec{x}_0) \land \trans(\vec{x}_0,\vec{x}_1) \limplies \inv(\vec{x}_1))  
\end{equation}
\noindent Each inductive invariant is a description of \emph{a}
fixed-point of the transition relation but is not
necessarily guaranteed to be \emph{the least} one, nor is it
guaranteed to include $\start(\vec{x})$ although many of the inductive
invariants we use will do.  For example, the
predicate $\true$ is an inductive invariant for all systems as it
describes the complete state space.
From an inductive invariant we can find loop invariants and function
and thread summaries by projecting on to a subset of variables $\vec{x}$.

Many verification tasks can be reduced to showing that the
reachable states do not intersect with a set of error states, 
denoted by the predicate $\err(\vec{x})$.  Techniques for proving
systems safe can be seen as computing an inductive invariant that is
disjoint from the error set.  Using existential second order
quantification (denoted $\secondorderexists$) we can formalise this as:
\begin{equation}
\begin{array}{rrl}
  \secondorderexists \inv \st & \forall \vec{x}_0, \vec{x}_1 \st & (\start(\vec{x}_0) \limplies \inv(\vec{x}_0)) \land \\
                &                 & (\inv(\vec{x}_0) \land \trans(\vec{x}_0,\vec{x}_1) \limplies \inv(\vec{x}_1)) \land \\
                &                 & (\inv(\vec{x}_0) \limplies \lnot \err(\vec{x}_0))
\end{array}
\end{equation}
Alternatively, if the system is not safe, then there is a reachable
error state.  One way of showing this is to find a concrete,
$n$-step counterexample\footnote{If the state space is finite and the system
is not safe there is necessarily a finite, concrete counterexample.
For infinite state spaces there are additional issues such as errors
only reachable via infinite counterexamples and which fixed-points
can be described by a finite formulae.}:
\begin{equation}
  \exists \vec{x}_0, \dots, \vec{x}_n \st \start(\vec{x}_0) \land
  \bigland{i \in [0,n-1]}  \trans(\vec{x}_i,\vec{x}_{i+1})
  \land \err(\vec{x}_n)
\end{equation}

\subsection{Existing Techniques}
\label{section:algorithm}

Viewing program verification as existential second-order logic allows
a range of existing tools to be characterised in a common framework
and thus compared and contrasted.  This section reviews some of the
more widely used approaches.  The following abbreviations,
corresponding to $k$ steps of the transition system and the first $k$
states being error free, will be used:
\begin{eqnarray*}
    \strans[k] = \bigland{i \in [0,k-1]}\trans(\vec{x}_i,\vec{x}_{i+1})
    & \hspace{1cm} &
    \sprop[k]  = \bigland{i \in [0,k-1]} \lnot \err(\vec{x}_i)
\end{eqnarray*}

\paragraph{Bounded Model Checking (BMC)}\cite{BCCZ99}  
focuses on refutation by picking a
\define{unwinding limit} $k$ and solving:
\begin{equation}
  \exists \vec{x}_0, \dots, \vec{x}_k \st \start(\vec{x}_0) \land
  \strans[k]
  \land
  \lnot \sprop[k+1]
\end{equation}
\noindent Models of this formula correspond to concrete
counterexamples of some length $n \lte k$.  The unwinding limit gives
an \emph{under-approximation} of the set of reachable states and thus
can fail to find counterexamples that take a large number of
transition steps.  In practice BMC works well as the formula is
existentially quantified and thus is in a fragment handled well by SAT
and SMT solvers.  There are also various simplifications that can
reduce the number of variables (see Section \ref{details:SSA}).

\paragraph{Incremental BMC (IBMC)} (e.g.~\cite{ES03b}) uses repeated BMC 
(often optimised
by using the solver incrementally) checks with increasing bounds to avoid the
need for a fixed bound.  If the bound starts at 0 (i.e.~checking
$\exists x_0 \st \start(\vec{x}_0) \land \err(\vec{x}_0)$) and is
increased linearly (this is the common use-case), then it can be
assumed that there are no errors at previous states, giving a simpler test:
\begin{equation}
  \exists \vec{x}_0, \dots, \vec{x}_k \st 
  \start(\vec{x}_0) \land 
  \strans[k] \land \sprop[k]
  \land
  \err(\vec{x}_k)
\end{equation}

\paragraph{K-Induction}\cite{SSS00} can be viewed as an extension of IBMC 
that can show system
safety as well as produce counterexamples.  It makes use of
\define{$k$-inductive invariants}, which are predicates that have the
following property:
\begin{equation}
  \forall \vec{x}_0 \dots \vec{x}_k \st \sinv[k] \land \strans[k] \limplies \kinv(\vec{x}_k)
\end{equation}
\noindent where
\begin{equation*}
  \sinv[k] = \bigland{i \in [0,k-1]} \kinv(\vec{x}_i)
\end{equation*}
\noindent $k$-inductive invariants have the following useful properties:
\begin{itemize}
\item{Any inductive invariant is a $1$-inductive invariant and vice versa.}
\item{Any $k$-inductive invariant is a $(k+1)$-inductive invariant.}
\item{A (finite) system is safe if and only if there is a $k$-inductive
  invariant $\kinv$ which satisfies:
\begin{equation}
\begin{array}{rrl}
   & \forall \vec{x}_0 \dots \vec{x}_k \st & 
    \left(\start(\vec{x}_0) \land \strans[k] \limplies \sinv[k] \right)
    \land \\
   &                 & 
    \left(\sinv[k] \land \strans[k] \limplies \kinv(\vec{x}_k)\right)
    \land \\
   &                 & 
     \left(\kinv(\vec{x}_k) \limplies \lnot \err(\vec{x}_k)\right)
\end{array}
\end{equation}
}
\end{itemize}
\noindent Showing that a $k$-inductive invariant exists is sufficient to
show that an inductive invariant exists \emph{but it does not imply
  that the $k$-inductive invariant is an inductive invariant}.  Often
the corresponding inductive invariant is significantly more complex.
Thus $k$-induction can be seen as a trade-off between 
invariant \emph{generation} and \emph{checking}
as it is a means to benefit as much as possible from 
simpler invariants by using a more complex property check.

Finding a candidate $k$-inductive invariant is hard so implementations
often use $\lnot \err(\vec{x})$. Similarly to IBMC, linearly
increasing $k$ can be used to simplify the expression by assuming
there are no errors at previous states:
\begin{equation}
\begin{array}{rrl}
 & \exists \vec{x}_0, \dots, \vec{x}_k \st &
  (\start(\vec{x}_0) \land \strans[k] \land \sprop[k] \land \err(\vec{x}_k)) \lor\\
 &                 & 
  (\strans[k] \land \sprop[k] \land \err(\vec{x}_k))
\end{array}
\end{equation}
\noindent A model of the first part of the disjunct is a concrete
counterexample ($k$-induction subsumes IBMC) and if the whole
formula has no models, then $\lnot \err(\vec{x})$ is a $k$-inductive
invariant and the system is safe.

\paragraph{Abstract Interpretation}\cite{CC77} While BMC and IBMC compute
under-approximations of the set of reachable states, the classical use
of abstract interpretation is to compute inductive invariants that
include $\start(\vec{x})$ and thus are over-approximations of the set
of reachable states.  Elements of an abstract domain can be understood
as sets or conjuncts of formulae \cite{DD13}, so abstract interpretation
can be seen as:
\begin{equation}
\begin{array}{rrl}
  \secondorderexists \ainv \in \adom \st & \forall \vec{x}, \vec{x}_1 \st & (\start(\vec{x}) \limplies \ainv(\vec{x})) \land \\
                &                 & (\ainv(\vec{x}) \land \trans(\vec{x},\vec{x}_1) \limplies \ainv(\vec{x}_1))
\end{array}
\end{equation}
\noindent where $\adom$ is the set of formulae described by 
the chosen abstract domain. As a second step then one checks:
\begin{equation}
\forall \vec{x} \st \ainv(\vec{x}) \limplies \lnot \err(\vec{x})  
\end{equation}
\noindent If this has no models then the system is safe, otherwise the
safety cannot be determined without finding a more restrictive $\ainv$
or increasing the set $\adom$, i.e. choosing a more expressive abstract
domain.  

\subsection{Our Algorithm: \algorithmName}

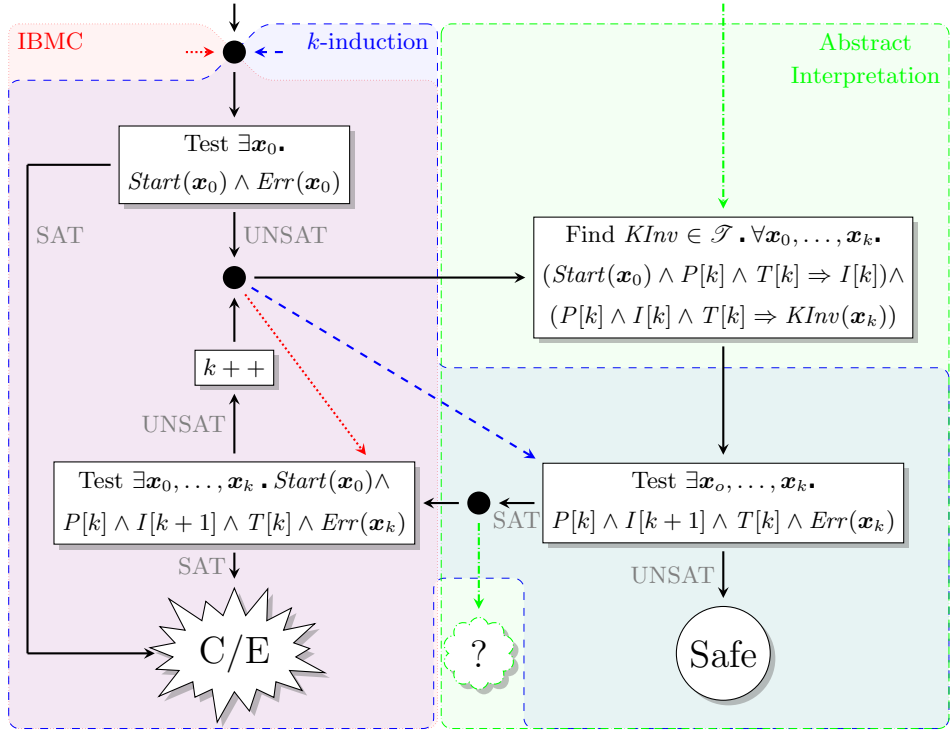
\begin{figure}[t]

\begin{center}
\begin{tikzpicture}

\tikzset{stage/.style = {drop shadow, draw, fill=white, shape=rectangle}}
\tikzset{final/.style = {stage}}
\tikzset{merge/.style = {fill=black, shape=circle}}

\tikzset{result/.style = {}}


\tikzset{ibmc/.style = {red, densely dotted}}
\tikzset{kind/.style = {blue, dashed}}
\tikzset{absint/.style = {green, densely dashdotted}}

\tikzset{blink/.style = {thick}}
\tikzset{hlink/.style = {blink, -stealth, shorten >=3pt}}
\tikzset{tlink/.style = {blink, shorten <=3pt}}
\tikzset{link/.style = {hlink, blink, tlink}}

\tikzset{abox/.style = {ultra thin, rounded corners, fill, fill opacity=0.05}}

\path
  let
   \n{answerline} = {2cm},
   \n{outsideoffset} = {2.75cm},
   \n{rectx} = {6.5cm},
   \n{recty} = {3cm}
  in
   coordinate (answerline) at (0,-1*\n{answerline})
   coordinate (outsidelow) at (-1*\n{outsideoffset}, -1*\n{answerline})
   coordinate (start) at (0, 2*\n{recty})
   coordinate (init) at (0, 1.5*\n{recty})
   coordinate (outsidehigh) at (outsidelow |- init)
   coordinate (initsplit) at (0, \n{recty})
   coordinate (inv) at (\n{rectx}, \n{recty})
   coordinate (invce) at (\n{rectx}, 0)
   coordinate (invcesplit) at (0.5*\n{rectx}, 0)
   coordinate (concretece) at (0,0)
   coordinate (increment) at (0,0.6*\n{recty})
   coordinate (ce) at (answerline -| concretece)
   coordinate (unknown) at (answerline -| invcesplit)
   coordinate (safe) at (answerline -| invce)
 ;

\path
  let
   \n{step} = {0.75cm}
  in
   (start)
   +( 0, 0.5*\n{step}) coordinate (boxtop)
   +( 0,-0.5*\n{step}) coordinate (boxtoplower)
   +( 0,  \n{step}) coordinate (kikistart)
   +(-0.5*\n{step},0) coordinate (ibmcstartpart)
   +(-1*\n{step},0) coordinate (ibmcstart)
   +( 0.5*\n{step},0) coordinate (kindstartpart)
   +( 1*\n{step},0) coordinate (kindstart)
   coordinate (absintstart) at (kikistart -| inv)
  ;


\path
  let
   \n{insetwidth} = {0.6cm},
   \n{insetheight} = {1cm}
  in
 (ce)
 +(0,-1) coordinate (boxbottom)
 (invce)
 +(3,0) coordinate (boxright)
 (concretece)
 +(-3,0) coordinate (boxleft)
 (unknown)
 +(-0.5,0) coordinate (innerleft)
 +(-1*\n{insetwidth},0) coordinate (insetleft)
 +( 1*\n{insetwidth},0) coordinate (insetright)
 +( 0, 1*\n{insetheight}) coordinate (insettop)
 (concretece)
 +(2.7,0) coordinate (innerright)
 ;

\draw [abox, ibmc] (boxleft |- boxtop) -- (ibmcstartpart |- boxtop) -- (kindstartpart |- boxtoplower) -- (boxtoplower -| innerright) -- (innerright |- boxbottom) -- (boxleft |- boxbottom) -- cycle;

\node [anchor=north west, ibmc] at (boxleft |- boxtop) {IBMC};

\draw [abox, kind] (innerright |- boxtop) -- (kindstartpart |- boxtop) -- (ibmcstartpart |- boxtoplower) -- (boxtoplower -| boxleft) -- (boxleft |- boxbottom) -- (insetleft |- boxbottom) -- (insetleft |- insettop) -- (insettop -| insetright) -- (insetright |- boxbottom) -- (boxbottom -| boxright) -- (boxright |- increment) -- (innerright |- increment) -- cycle;

\node [anchor=north east, kind] at (innerright |- boxtop) {$k$-induction};

\draw [abox, absint] (boxtop -| innerleft) -- (innerleft |- boxbottom) -- (boxright |- boxbottom) -- (boxright |- boxtop) -- cycle;

\node [anchor=north east, absint, rectangle split, rectangle split parts=2, rectangle split draw splits=false] at (boxright |- boxtop) {Abstract \nodepart{second} Interpretation};

\node [merge] (Start) at (start) {};
\node [stage, rectangle split, rectangle split parts=2, rectangle split draw splits=false] (Init) at (init) {Test $\exists \vec{x}_0 \st$ \nodepart{second} $\sstart(\vec{x}_0) \land \serr(\vec{x}_0)$};

\node [merge] (Initsplit) at (initsplit) {};

\node [stage, rectangle split, rectangle split parts=3, rectangle split draw splits=false] (Inv) at (inv) {Find $\skinv \in \tdom \st \forall \vec{x}_0 , \dots, \vec{x}_k \st$ \nodepart{second}
$(\sstart(\vec{x}_0) \land \sprop[k] \land \strans[k] \limplies \sinv[k]) \land$ \nodepart{third}
$(\sprop[k] \land \sinv[k] \land \strans[k] \limplies \skinv(\vec{x}_k))$};

\node [stage, rectangle split, rectangle split parts=2, rectangle split draw splits=false] (Invce) at (invce) 
{Test $\exists \vec{x}_o , \dots, \vec{x}_k \st$ \nodepart{second} $\sprop[k] \land \sinv[k+1] \land \strans[k] \land \serr(\vec{x}_k)$};
\node [merge] (Invcesplit) at (invcesplit) {};

\node [stage, rectangle split, rectangle split parts=2, rectangle split draw splits=false] (Concretece) at (concretece)
{Test $\exists \vec{x}_0 , \dots, \vec{x}_k \st \sstart(\vec{x}_0) \land$ \nodepart{second} $\sprop[k] \land \sinv[k+1] \land \strans[k] \land \serr(\vec{x}_k)$};
\node [stage] (Increment) at (increment) {$k++$};

\node [final, starburst] (CE) at (ce) {\Large C/E};
\node [final, cloud, absint, fill=white] (Unknown) at (unknown) {\Large \color{black} ?};
\node [final, circle] (Safe) at (safe) {\Large Safe};


\draw [link] (kikistart) -- (Start.north);

\draw [link] (Start.south) -- (Init.north);
\draw [link] (Init.south) -- (Initsplit.north);
\node [result, right] (UNSAT) at ($(Init.south)!0.5!(Initsplit.north)$) {\result{UNSAT}};

\draw [link] (Initsplit.east) -- (Inv.west);

\draw [link] (Inv.south) -- (Invce.north);
\draw [link] (Invce.west) -- (Invcesplit.east);
\node [result, below] at ($(Invce.west)!0.5!(Invcesplit.east)$) {\result{SAT}};

\draw [link] (Invcesplit.west) -- (Concretece.east);

\draw [link] (Concretece.north) -- (Increment.south);
\node [result, left] at ($(Concretece.north)!0.5!(Increment.south)$) {\result{UNSAT}};
\draw [link] (Increment.north) -- (Initsplit.south);

\draw [link] (Invce.south) -- (Safe.north);
\node [result, left] at ($(Invce.south)!0.5!(Safe.north)$) {\result{UNSAT}};
\draw [link] (Concretece.south) -- (CE.north);
\node [result, left] at ($(Concretece.south)!0.5!(CE.north)$) {\result{SAT}};

\draw [tlink] (Init.west) -- (outsidehigh);
\draw [blink] (outsidehigh) -- (outsidelow);
\draw [hlink] (outsidelow) -- (CE.west);
\node [result, right] at (outsidehigh |- UNSAT) {\result{SAT}};

\draw [ibmc, link] (ibmcstart) -- (Start);
\draw [ibmc, link] (Initsplit.-60) -- (Concretece.17);

\draw [kind, link] (kindstart) -- (Start);
\draw [kind, link] (Initsplit.-30) -- (Invce.north west);

\draw [absint, link] (absintstart) -- (Inv.north);
\draw [absint, link] (Invcesplit.south) -- (Unknown);

\end{tikzpicture}
\end{center}
  
  \caption{The \algorithmName\ algorithm (colours in online version)}
  \label{figure:kiki-flow-chart}
\end{figure}

The phases of the \algorithmName\ algorithm are presented as a flow chart in
Figure~\ref{figure:kiki-flow-chart} with black arrows denoting 
transitions.  Initially, $k = 1$ and $\tdom$ is
a set of predicates that can be used as invariant with $\top \in
\tdom$ (see Section~\ref{section:algorithm-details} for details
of how this is implemented).

After an initial test to see if any start states are
errors\footnote{If the transition system is derived from software and
  the errors are generated from assertions this will be impossible and
the check can be skipped.}, \algorithmName\ computes a $k$-inductive
invariant that covers the initial state and includes the assumption
that there are no errors in earlier states.
  The invariant is then checked to see whether it is 
sufficient to show safety.  If there are possible reachable error
states then a second check is needed to see if the error is reachable
in $k$ steps (a genuine counterexample) or whether it is a potential
artefact of a too weak invariant.  In the latter case, $k$ is incremented
so that a stronger ($k$-)invariant can be found and the algorithm loops.

Also displayed in Figure \ref{figure:kiki-flow-chart} are the steps of
incremental BMC, $k$-induction and classical over-approximating abstract
interpretation, given, respectively by the red dotted, blue dashed and
green dashed/dotted boxes and arrows.
  \algorithmName\ can simulate $k$-induction by having
$\tdom = \{ \top \}$ and incremental BMC by over-approximating
the first SAT check.  Classical over-approximate abstract
interpretation can be simulated by having $\tdom = \adom$
and terminating with the result ``unknown'' if the first SAT check
finds a model.  These simulations give an intuition for the proof of
the following results:

\begin{theorem}~\\
  \begin{compactitem}
  \item{When \algorithmName\ terminates it gives either a
    $k$-inductive invariant sufficient to show safety or a length $k$
    counterexample.}

  \item{If IBMC or $k$-induction terminate with a length $k$ counterexample, then
    \algorithmName\ will terminate with a length $k$ counterexample.}

  \item{If $k$-induction terminates with a $k$-inductive invariant
    sufficient to show safety, then \algorithmName\ will terminate with
    a $k$-inductive invariant sufficient to show safety.}

  \item{If an (over-approximating) abstract interpreter returns an
    inductive invariant $\ainv$ that is sufficient to show safety and
    $\adom \subseteq \tdom$, then \algorithmName\ will
    terminate with $k = 1$ and an inductive invariant sufficient to
    show safety.
    }
  \end{compactitem}
\end{theorem}

Hence \algorithmName\ strictly generalises its components by
exploiting the following synergies between them: unrolling $k$ times
helps abstract interpretation to generate stronger invariants, namely
$k$-invariants, which are further strengthened by the additional facts
known from not having found a counterexample for $k-1$ iterations;
stronger invariants help $k$-induction to successfully prove
properties more often; and constraining the state space by invariants
ultimately accelerates the countermodel search in BMC.  We will
observe these synergies also experimentally in
Section~\ref{section:experiments}.

\section{Algorithm Details}
\label{section:algorithm-details}

Section \ref{section:algorithm-concepts} introduced
\algorithmName\ but omitted a number of details which are important
for implementing the algorithm efficiently.  Key amongst these are the
encoding from program to transition system and the generation of
$k$-inductive invariants.

\begin{figure}[t]
\centering
  \begin{subfigure}[t]{0.20\textwidth}
    \centering
{
\scriptsize
\begin{lstlisting}
void main()
{
  unsigned x = 0;


  while (x<10)
  {
    ++x;
  }

  assert(x==10);
}
\end{lstlisting}
}

\caption{The program}
\label{figure:program}
  \end{subfigure}
\hspace{1em} 
  \begin{subfigure}[t]{0.4\textwidth}

{
\scriptsize
\begin{lstlisting}

guard#0 == TRUE
x#0 == 0u

guard#1 == guard#0
x#phi1 == (guard#ls0 ? x#lb1 : x#0)
guard#2 == (x#phi1 < 10) && guard#1
x#2 == 1u + x#phi1

guard#3 == !(x#phi1 < 10) && guard#1
x#phi1 == 10u || !guard#3

\end{lstlisting}

\vspace{8pt} 
}

\caption{The annotated SSA}
\label{figure:ssa}
  \end{subfigure}

  \caption{Conversion from program to SSA}
  \label{figure:program-to-SSA}
\end{figure}

\subsection{SSA Encoding}
\label{details:SSA}

The presentation of \algorithmName\ used transition systems and it
is possible to implement this directly.
However the symbolic transition systems generated by software have
structural properties that can be exploited.
In most states the value of the program
counter uniquely identifies its next value
(i.e.~most instructions do not branch) and most transitions update a
single variable.  Thus states in the transition can be merged by
substituting in the symbolic values of updated variables, so reducing
the size of the formulae generated.

Rather than building the transition system and then reducing it, it
is equivalent and more efficient to convert the program to \emph{single static
 assignment form} (SSA).
For acyclic code, the SSA is a formula that exactly represents
the strongest post condition of running the code and generation of
this is a standard technique found in most software BMC and Symbolic
Execution tools.
We extend this with an over-approximate conversion of loops so that
the SSA allows us to reason about abstractions of a program with a solver.

\begin{figure*}[t!]
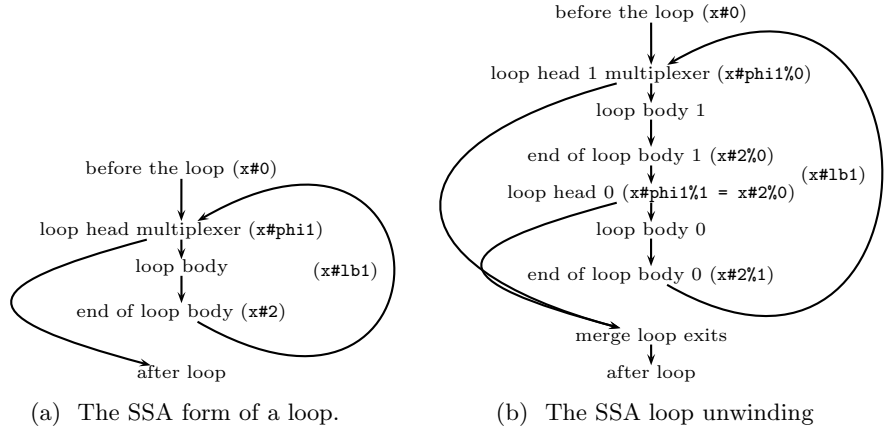

    \centering
    \begin{subfigure}[b]{0.5\textwidth}
        \centering \scriptsize
\begin{tabular}{c}
\rnode{i}{before the loop (\texttt{x\#0})} \\[5ex]
\rnode{h}{loop head multiplexer (\texttt{x\#phi1})} \\[2ex]
\rnode{b}{loop body} \\[3ex]
\rnode{e}{end of loop body (\texttt{x\#2})} \\[5ex]
\rnode{a}{after loop}
\end{tabular}
\ncline[arrows=->]{i}{h}
\ncline[arrows=->]{h}{b}
\ncline[arrows=->]{b}{e}
\nccurve[angleA=-30,angleB=30,arrows=->,ncurv=6]{e}{h}\naput{(\texttt{x\#lb1})}
\nccurve[angleA=-165,angleB=165,arrows=->,ncurv=3]{h}{a}
\caption{\label{fig:loopssa}
The SSA form of a loop.
}
    \end{subfigure}%
    ~ 
    \begin{subfigure}[b]{0.5\textwidth}
        \centering \scriptsize
\begin{tabular}{c}
\rnode{i}{before the loop (\texttt{x\#0})} \\[5ex]
\rnode{h1}{loop head 1 multiplexer (\texttt{x\#phi1\%0})} \\[2ex]
\rnode{b1}{loop body 1} \\[3ex]
\rnode{e1}{end of loop body 1 (\texttt{x\#2\%0})} \\[2ex]
\rnode{h0}{loop head 0 (\texttt{x\#phi1\%1 = x\#2\%0})} \\[2ex]
\rnode{b0}{loop body 0} \\[3ex]
\rnode{e0}{end of loop body 0 (\texttt{x\#2\%1})} \\[5ex]
\rnode{a}{merge loop exits}\\[2ex]
\rnode{aa}{after loop}
\end{tabular}
\ncline[arrows=->]{i}{h1}
\ncline[arrows=->]{h1}{b1}
\ncline[arrows=->]{b1}{e1}
\ncline[arrows=->]{e1}{h0}
\ncline[arrows=->]{h0}{b0}
\ncline[arrows=->]{b0}{e0}
\ncline[arrows=->]{a}{aa}
\nccurve[angleA=-30,angleB=30,arrows=->,ncurv=3]{e0}{h1}\naput{(\texttt{x\#lb1})}
\nccurve[angleA=-165,angleB=165,arrows=->,ncurv=3]{h0}{a}
\nccurve[angleA=-165,angleB=165,arrows=->,ncurv=2]{h1}{a}
\caption{\label{fig:loopunwinding}
The SSA loop unwinding
}
    \end{subfigure}
    \caption{Illustrations of various SSA encodings}
\end{figure*}

Figure \ref{figure:program-to-SSA} gives an example of the conversion.
The SSA has been made acyclic by cutting loops at the end of the loop
body: the variable\footnote{Variable name suffixes are use to denote
  the multiple \emph{logical} variables that correspond to a single
  \emph{program} variable at different points in the execution.}
 \texttt{x\#2} at the end of the loop body
(``poststate'') corresponds to \texttt{x\#lb1}, which is
fed back into the loop head (``prestate'').
A non-deterministic choice (using the free Boolean variable
\texttt{guard\#ls0}) is introduced at the loop head in order to join the
values coming from before the loop and from the end of the loop body.
Figure~\ref{fig:loopssa} illustrates how the SSA statements express
control flow.

It is easy to see that this representation ``havocs'' loops because
\texttt{x\#lb1} is a free variable -- this is why its models are an
over-approximation of actual program traces.  Precision can be
improved by constraining the feedback variable \texttt{x\#lb1} by
means of a \emph{loop invariant} which we are going to infer.  Any
property that holds at loop entry (\texttt{x\#0}) and at the end of
the body (\texttt{x\#2}) can then be assumed to hold on the feedback
variable \texttt{x\#lb1}.

Loop unwinding is performed in the usual fashion; the conversion to SSA
simply repeats the conversion of the body of the loop.  Figure
\ref{fig:loopunwinding} illustrates an example of this.  The top-most
loop head multiplexer is kept and its feedback variable is constrained
with the bottom-most loop unwinding.  The only subtlety is that the value of
variables from different loop exits must be merged.  This can be
achieved by use of the \texttt{guard} variables which track the
reachability of various program points for a given set of values.
The unwinding that we perform is incremental, in the sense that the
construction of the formula is monotonic.  Assumptions have to be used
to deal with the end of loop merges as there always has to be a
case for ``value is merged from an unwinding that has not been added yet''
and this has to be assumed false.

A more significant example is given in\pponly{ the extended version \cite{extended-version}}\rronly{ Appendix \ref{sec:running}}.

\subsection{Invariant Inference via Templates}
\label{section:invariantsviatemplates}

A key phase of \algorithmName\ is the generation of $\skinv$, a
k-inductive invariant.
 Perhaps the most obvious approach is to use an off-the-shelf abstract
 interpreter.  This works but will fail to exploit the real power of
 \algorithmName.  Each iteration, \algorithmName\ unrolls loops one
 more step (which can improve the invariant given by an abstract
 interpreter) and adds assumptions that previous unwindings do not
 give errors.  Without backwards propagation it is difficult for an
 abstract interpreter to make significant use of these assumptions.
 For example, an abstract interpretation with intervals would need
 backwards propagation to make use of \texttt{assume(x + y < 10)}.
 Thus we use a solver-based approach to computing $\skinv$ as it can
 elegantly exploit the assumptions that are added without needing to
 (directly) implement transformers.

Directly using a solver we would need to handle (the
existential fragment of) second-order logic.
As these are not currently available, we reduce to a 
problem that can be solved by iterative application of a first-order
solver.
We restrict ourselves to finding invariants $\kinv$ of the form
$\templ(\vec{x},\vec{\delta})$ where $\templ$ is a fixed expression, a
so-called \emph{template}, over program variables $\vec{x}$ and
template parameters $\vec{\delta}$ (see Section~\ref{sec:domains}).
This restriction is analogous to choosing an abstract domain in an
abstract interpreter and has similar effect -- $\tdom$ only contains a
the formulae that can be described by the template.
Fixing a template reduces the second-order search for an invariant to
the first-order search for template \emph{parameters}:
\begin{equation}\label{equ:invtempl}
\begin{array}{rrl}
 \exists \vec{\delta} \st  & \forall \vec{x}_0 \dots \vec{x}_k \st & 
    \left(\start(\vec{x}_0) \land \strans[k] \limplies 
          \templ[k](\vec{\delta}) \right)
    \land \\
   &                 & 
    \left(\templ[k](\vec{\delta})
          \land \strans[k] \limplies \templ(\vec{x_k},\vec{\delta}) \right)
\end{array}
\end{equation}
with
$\templ[k](\vec{\delta}) = \bigland{i \in [0,k-1]} \templ(\vec{x_i},\vec{\delta})$.
Although the problem is now expressible in first-order logic, it
contains quantifier alternation which poses a problem for current SMT
solvers.
However, we can solve this problem by iteratively checking 
the negated formula (to turn $\forall$ into $\exists$)
for different choices of constants $\vec{d}$ for the parameters
$\vec{\delta}$; as for the second conjunct in
(\ref{equ:invtempl}):

\begin{equation}\label{equ:qfinvtempl}
\exists \vec{x}_0 \dots \vec{x}_k \st \neg \big(\templ[k](\vec{d}) \wedge
\strans[k] \limplies \templ(\vec{x_k},\vec{d})\big)
\end{equation}

The resulting formula can be expressed in quantifier-free logics and
efficiently solved by SMT solvers.  Using this as a building block,
one can solve this $\exists\forall$ problem (see Section \ref{sec:qe}).

\subsection{Guarded Template Domains}\label{sec:domains}

As discussed in the previous section, we use templates and repeated
calls (with quantifier-free formulae) to a first-order solver to
compute $k$-inductive invariants.

An abstract value $\vec{d}$ represents, i.e. \emph{concretises} to,
the set of all $\vec{x}$ that satisfy the formula
$\templ(\vec{x},\vec{d})$.
We require an abstract value $\bot$ denoting the empty set
$\templ(\vec{x},\bot) \equiv \false$, and $\top$ for
the whole domain of~$\vec{x}$: $\templ(\vec{x},\top) \equiv \true$.

\paragraph{Template polyhedra}
We use template polyhedra \cite{SSM05},
a class of templates for numerical variables which have the form
$\templ = (\mathbf{A}\vec{x}\leq\vec{\delta})$ where
$\mathbf{A}$ is a matrix with fixed coefficients.
Subclasses of such templates include \emph{Intervals},
which require constraints $\vecv{1}{-1}x_i\leq\vecv{\delta_{i1}}{\delta_{i2}}$
for each variable $x_i$, \emph{Zones} (differences), and
\emph{Octagons} \cite{Min01a}.
The $r^{\textit{th}}$ \emph{row} of the template are the constraint
generated by the $r^{\text{th}}$ row of matrix $\mathbf{A}$.

In our template expressions, variables $\vec{x}$ are
\emph{bit-vectors} representing signed or unsigned integers. 
These variables can be mixed in template constraints. Type promotion
rules are applied such that the bit-width of the types of the
expressions are extended in order to avoid arithmetic under- and
overflows in the template expressions.
$\top$ corresponds to the respective maximum values in the promoted
type, whereas $\bot$ must be encoded as a special symbol.

\paragraph{Guarded templates}
Since we use SSA form
rather than
control flow graphs, we cannot use numerical templates directly.
Instead we use \emph{guarded templates}.
In a guarded template each row $r$ is of the form
$G_r \limplies \basetempl_r$ for the $r^{\text{th}}$ row $\basetempl_r$ of the
base template domain (e.g. template polyhedra).
$G_r$ is the conjunction of the SSA guards $g_i$
associated with the definition of variables $x_i$ occurring in $\basetempl_r$.
$G_r$ denotes the guard associated to variables $\vec{x}$ appearing at the
loop head, and $G_r'$ the guard associated to the variables
$\vec{x}'$ at the end of the respective loop body.
Hence, template rows for different loops have different guards.

A guarded template in terms of the variables at the loop
head is hence of the form $\templ(\vec{x}_0,\vec{\delta}) = \bigwedge_r
G_r(\vec{x}_0) \limplies \basetempl_r(\vec{x}_0,\vec{\delta})$.
Replacing parameters  $\vec{\delta}$ by the values $\vec{d}$ 
we get the invariants $\templ(\vec{x},\vec{d})$ at the loop heads.

For the example program in Section~\ref{details:SSA}, we have the following
guarded interval template:
$$
\templ(\texttt{x\#lb1},(\delta_1,\delta_2)) 
= \left\{
\begin{array}{lcr}
\texttt{guard\#1} \wedge \texttt{guard\#ls0} &\limplies & \texttt{x\#lb1} \leq \delta_1 \\
\texttt{guard\#1} \wedge \texttt{guard\#ls0} &\limplies & -\texttt{x\#lb1} \leq \delta_2 \\
\end{array}\right.
$$ 

We denote $\templ'(\vec{x}_1,\vec{\delta}) = \bigwedge_r G_r'(\vec{x}_1) \limplies
\basetempl_r(\vec{x}_1,\vec{\delta})$ the guarded template expressed in terms of the variables at the end of the loop body. Here, we have to  express the join of the initial value at the loop head (like \texttt{x\#0}) and the values that are fed back into the loop head (like \texttt{x\#2}). For the example above, the corresponding guarded template is as follows:
$$
\templ'(\texttt{x\#2},(\delta_1,\delta_2)) 
= \left\{
\begin{array}{l}
(\mathit{pg} \Leftrightarrow \texttt{guard\#2}) ~\wedge~
(\mathit{ig} \Leftrightarrow \texttt{guard\#1} \wedge \neg\texttt{guard\#ls0}) \wedge \\
((\mathit{ig} \Rightarrow x'=\texttt{x\#0}) \wedge 
   (\mathit{pg} \wedge \neg\mathit{ig} \Rightarrow x'=\texttt{x\#2})) \wedge \\
(\mathit{pg} \vee \mathit{ig} \Rightarrow x' \leq \delta_1) \wedge 
(\mathit{pg} \vee \mathit{ig} \Rightarrow -x' \leq \delta_2)
\end{array}\right.
$$ 

\subsection{Accelerated Solving of the $\exists\forall$ Problem}
\label{sec:qe}

As discussed in Section \ref{section:invariantsviatemplates}, it is
necessary to solve an $\exists\forall$ problem to find values for
template parameters $\vec{\delta}$ to infer invariants.

\paragraph{Model enumeration.}
The well-known method \cite{RSY04,BKK11} for solving this problem 
in formula (\ref{equ:qfinvtempl})
using SMT solvers repeatedly checks satisfiability of the formula for
an abstract value $\vec{d}$ (starting with $\vec{d}=\bot$):
\begin{equation}\label{equ:indcheck}
\templ[k](\vec{d}) \land \strans[k] \wedge \neg \templ'(\vec{x_k},\vec{d})
\end{equation}
If it is unsatisfiable, then we have found an
invariant; otherwise we join the model returned by the solver with the
previous abstract value $\vec{d}$.

However, this method corresponds to performing a classical 
Kleene iteration on the abstract lattice up to convergence.
Convergence is guaranteed because our abstract domains are
finite. Though, the height of the lattice is enormous and even for a
one loop program incrementing an unconstrained 64-bit integer variable
the na\"ive algorithm will not terminate within human life time.
Hence, we are not going to use this method.

\paragraph{Optimisation.}
What we need is a convergence acceleration that 
makes the computational effort \emph{independent} from the number
of states and loop iterations.
To this end, we use a technique that is inspired by an encoding 
used by max-\emph{strategy iteration} methods
\cite{GS07,GM11,MS14b}.
These methods state the invariant inference problem over template
polyhedra as a disjunctive linear optimisation problem,
which is solved iteratively by an upward iteration in the lattice of
template polyhedra:
using SMT solving, a conjunctive subsystem (``strategy'') whose solution
extends the current invariant candidate is selected.  This subsystem is
then solved by an LP solver; the procedure terminates as soon as an
inductive invariant is found.

This method can only be used if the domain is convex and the parameter
values are ordered and monotonic w.r.t. concretisation, which holds
true, for example, 
for template polyhedra $\mathbf{A}\vec{x}\leq\vec{d}$ where $\vec{d}$
is a parameter, but not for those where $\mathbf{A}$ is a parameter.
If the operations in the transition relation satisfy certain
properties such as monotonicity of condition predicates, then the
obtained result is the least fixed point, i.e. the \emph{same} result
as the one returned by the na\"ive model enumeration above, but much
faster on average.

\paragraph{Our algorithm.}
We adapt this method to our setting with bit-vector variables and
guarded templates. Since we deal with finite domains (bit-vectors) we
can use \emph{binary search} as optimisation method instead of an LP solver.

The algorithm proceeds as follows:
We start by checking whether the current abstract value $\vec{d}$
(starting from $\vec{d}=\bot$) is inductive
(Equ.~(\ref{equ:indcheck})). If so, we have found an invariant;
otherwise there are template rows $R$ whose values are not inductive
yet.
We construct the system
\begin{equation}\label{equ:symbsys}
\hspace*{-3em}\bigland{i \in [0,k-1]} \left\{\begin{array}{rll}
& \bigwedge_{r\notin R} & G_r(\vec{x}_i) \limplies (\rowexpr_r(\vec{x}_i)\leq d_r) \\[1ex]
\wedge & \bigwedge_{r\in R} & G_r(\vec{x}_i) \limplies (\rowexpr_r(\vec{x}_i)\leq \delta_r)\\[1.5ex]
\end{array}\right\} 
\wedge \strans[k]
\wedge \bigwedge_{r\in R} G'_r(\vec{x}_k) \wedge (\delta_r\leq \rowexpr_r(\vec{x}_k))
\end{equation}
where $\rowexpr_r$ is the left-hand side of the inequality
corresponding to the $r^\text{th}$ row of the template.
Then we start the binary search for the optimal value of $\sum_{r\in R}\delta_r$ 
over this system.
The initial bounds for $\sum_{r\in R}\delta_r$ are as follows:
\begin{itemize}
\item The lower bound $\ell$ is $\sum_{r\in R} d'_r$ where $d'_r$ is the
  value of $\rowexpr_r(\vec{x}_k)$ in the model of the inductivity check
  (\ref{equ:indcheck}) above;
\item The upper bound $u$ is $\sum_{r\in R}\mathit{max\_value}(r)$ where
  $\mathit{max\_value}$ returns the maximum value that $\rowexpr_r(\vec{x}_k)$
  may have (dependent on variable type).
\end{itemize}

The binary search is performed by iteratively checking
(\ref{equ:symbsys}) for satisfiability under the assumption
$\sum_{r\in R}\delta_r\geq m$ where $m = \mathit{median}(\ell,u)$.  If
satisfiable, set $\ell := m$, otherwise set $u := m$ and repeat until
$\ell = u$.
The values of $\delta_r$ in the last satisfiable query are assigned to 
$d_r$ to obtain the new abstract value.
The procedure is then repeated by testing whether $\vec{d}$ is
inductive (\ref{equ:indcheck}).
Note that this algorithm uses a similar encoding for bound
optimisation as strategy iteration, but potentially requires a higher
number of iterations than strategy iteration.  This choice has been
made deliberately in order to keep the size of the generated SMT
formulas small, at the cost of a potentially increased number of
iterations.

A worked example is given in\pponly{ the extended version
  \cite{extended-version}}\rronly{ Appendix \ref{sec:running}}.

\section{Implementation}
\label{section:implementation}

We implemented \algorithmName\ in {\systemName},%
\footnote{Version 0.2. The source code of the tool and instructions
  for its usage can be found on
  \url{http://www.cprover.org/wiki/doku.php?id=2ls_for_program_analysis}. In the experiments we ran it with the option \texttt{-{}-competition-mode}.}
a verification tool built on the CPROVER framework, using
MiniSAT-2.2.0 as a back-end solver (although other SAT and SMT solvers
with incremental solving support can also be used).
\systemName\ currently inlines all functions when running
{\algorithmName}. 
The techniques described in Section \ref{section:algorithm-details}
enable a single solver instance to be used where constraints and
unwindings are added incrementally. 
This is essential because {\algorithmName} makes thousands of solver
calls for invariant inference and property checks.

Our implementation is generic w.r.t.\ matrix $\textbf{A}$ of the
template polyhedral domain. In our experiments, we observed that very
simple matrices $\textbf{A}$ generating interval invariants are
sufficient to compete with other state-of-the-art tools.

The tool can handle unrestricted sequential C programs (with the
exception of programs with irreducible control flow).  However,
currently, invariants are not inferred over array contents or
dynamically allocated data structures.

\section{Experiments}
\label{section:experiments}

We performed a number of experiments to demonstrate the utility and
applicability of {\algorithmName}.  All experiments were performed
on an Intel Xeon X5667 at 3\,GHz running Fedora 20 with 64-bit binaries.
Each individual run was limited to 13\,GB 
of memory and 900 seconds of
CPU time, 
enforced by the operating
system kernel.  We took the \emph{loops} meta-category (143 benchmarks) from the SV-COMP'15 benchmark
set.%
\footnote{\url{http://sv-comp.sosy-lab.org/2015/benchmarks.php}}

\subsection{\algorithmName\ Verifies More Programs Than the Algorithms it Simulates}

Table~\ref{tab:results} gives a comparison between \systemName\ running
\algorithmName\ (column 6) and \emph{the same system} running as an incremental
bounded model checker (IBMC) (column 2), incremental $k$-induction (i.e. without invariant inference, column 3)
and as an abstract interpreter (AI) (column 4).
 \algorithmName\ is more complete than each of the restricted modes. 
This is not self-evident since it could be much less efficient and,
thus, fail to solve the problems within the given time or memory limits.
$k$-induction can solve 60.8\% of the benchmarks, 13 more than IBMC. 
32\% of the benchmarks can be solved by abstract interpretation (bugs
are only exposed if they are reachable with 0 loop unwindings).  
\algorithmName\ solves 62.9\% of the benchmarks, 
proving 3 more properties than $k$-induction.

\begin{table}
\centering
\begin{tabular}{|l|ccc|c|c|c|c|}
\hline
& IBMC & $k$-induction & AI & portfolio & ~\algorithmName\ ~ &
                                                               CPAchecker & ESBMC \\
\hline
counterexamples & \bf 38 & \bf 38 &  17 & \bf 38 & \bf 38 & 36 &  35 \\
proofs          &    36 &  49 &  30 &  51 &  52 & 59 & \bf 91 \\
false proofs    &     0 &   0 &   0 &   0 &   0 &  2 &  12\\
false alarms    &     2 &   2 &   0 &   2 &   2 &  2 &   0\\
inconclusive    &     0 &   0 &  93 &   0 &   0 &  4 &   2\\
timeout         &    65 &  53 &   3 &  50 &  51 & 38 &   2 \\
memory out      &     2 &   1 &   0 &   2 &   0 &  2 &   1 \\
total runtime   & 17.1h & 13.8h & 0.89h & 13.3h & 13.2h & 10.9h & 0.54h \\
\hline
\end{tabular}~\\[1ex]
\caption{Comparison between {\algorithmName}, the algorithms it subsumes,
the portfolio, and CPAchecker. The rows false alarms and false proofs 
indicate soundness bugs of the tool implementations.
\label{tab:results}}
\end{table}

\subsection{\algorithmName\ is at Least as Good as Their Na\"ive Portfolio}

To show
that \algorithmName\ is more than a mixture of three techniques and
that they strengthen each other, consider column 5 of
Table~\ref{tab:results}.  This gives the results of an ideal portfolio
in which the three restricted techniques are run in parallel on and
the portfolio terminates when the first returns a conclusive result.
Thus the CPU time taken is three times the time taken by the fastest
technique for each benchmark (in practice these could be run in
parallel, giving a lower \emph{wall clock} time).
In our setup, \algorithmName\ had a disadvantage as each component of
virtual portfolio had the same 
memory restriction as
{\algorithmName}, thus effectively giving the portfolio three times as much memory.

Still, {\algorithmName} is slightly faster and more
accurate than the portfolio as can be seen in Table \ref{tab:results}.
The scatter plot in Figure~\ref{fig:results}a shows the results for
each benchmark:
one can observe that {\algorithmName} is up to one order of magnitude
slower on many unsafe benchmarks, which is obviously due to the
additional work of invariant inference that
{\algorithmName} has to perform in contrast to IBMC.
However, note that {\algorithmName} is faster than the 
portfolio on some safe and even one unsafe benchmarks.
This suggests that {\algorithmName} is more than the sum of its parts.

\subsection{\algorithmName\ is Comparable with State-of-the-Art Approaches}

\begin{figure}[t]
\begin{tabular}{@{\hspace{-1.5em}}c@{\hspace{1em}}c}
\begin{tikzpicture}[scale=0.75]
	\begin{loglogaxis} [xmin=.1,xmax=1500, ymin=.1, ymax=1500, xlabel=kIkI (time in seconds),
			ylabel=portfolio (time in seconds), 
			legend style={at={(0.8,0.15)},
			anchor=north,legend columns=-1 },
			]
\addplot [mark size=1pt,only marks,scatter,point meta=explicit symbolic,
	scatter/classes={s={mark=square},u={mark=triangle*,blue}},] 
	table [meta=label] {scatter-kiki-vbs-sas.dat};
	\legend{Safe,Unsafe}
\addplot [domain=.1:1500] {x};
\addplot [red,sharp plot, domain=.1:1500] {900}
          node [below] at (axis cs:10,850) {timeout};
\addplot [red,sharp plot, domain=.1:1500] coordinates{(900,.1) (900,1500)}
          node [left,rotate=90] at (axis cs:700,10) {timeout}
 node [right,black] at (axis cs:10,3) {portfolio faster}
 node [right,black] at (axis cs:1,55) {kIkI faster};
\end{loglogaxis}
\end{tikzpicture}
 &
\begin{tikzpicture}[scale=0.75]
	\begin{loglogaxis} [xmin=.1,xmax=1500, ymin=.1, ymax=1500, xlabel=kIkI (time in seconds),
			ylabel=CPAchecker (time in seconds),
			legend style={at={(0.8,0.15)},
			anchor=north,legend columns=-1 },
			]
\addplot [mark size=1pt,only marks,scatter,point meta=explicit symbolic,
	scatter/classes={s={mark=square},u={mark=triangle*,blue}},] 
	table [meta=label] {scatter-kiki-cc-sas.dat};
	\legend{Safe,Unsafe}
\addplot [domain=.1:1500] {x};
\addplot [red,sharp plot, domain=.1:1500] {900}
          node [below] at (axis cs:10,850) {timeout};
\addplot [red,sharp plot, domain=.1:1500] coordinates{(900,.1) (900,1500)}
          node [left,rotate=90] at (axis cs:700,150) {timeout}
 node [right,black] at (axis cs:10,3) {CPAchecker faster}
 node [right,black] at (axis cs:1,55) {kIkI faster};
\end{loglogaxis}
\end{tikzpicture} \\
(a) & (b)
\end{tabular}
\caption{\label{fig:results}
Runtime Comparison
}
\end{figure}
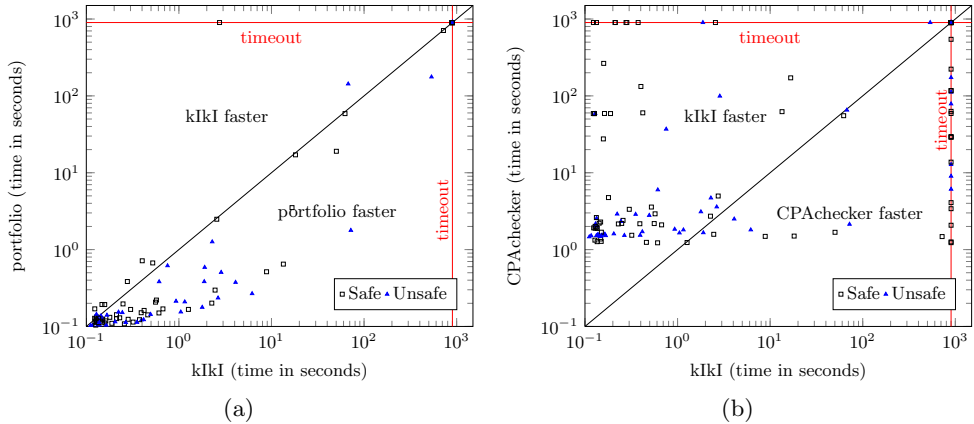

We compared our implementation of \algorithmName\ with
CPAchecker%
\footnote{SVCOMP'15 version, http://cpachecker.sosy-lab.org/}%
, and ESBMC%
\footnote{SVCOMP'15 version, http://www.esbmc.org/}%
, which uses $k$-induction.
The results are shown in the last three columns in Table \ref{tab:results}
and in the scatter plot in Figure~\ref{fig:results}b. Additional
results are given in \pponly{the extended version
  \cite{extended-version}}\rronly{Appendix \ref{sec:further}}.
In comparison to CPAchecker, the winner of SVCOMP'15,
our prototype of {\algorithmName} is overall a bit slower and proves 
fewer properties (due to more timeouts), but as
Figure~\ref{fig:results}b shows, it significantly outperforms
CPAchecker on most benchmarks.
ESBMC exposes fewer bugs, but proves many more properties and is 
significantly faster. However, it has 6 times
more soundness bugs than our implementation.%
\footnote{The two false alarms in our current implementation are due to 
limited support for dynamic memory allocation.}
These results show that our prototype implementation of
\algorithmName\ can keep up with state-of-the-art verification tools.

\section{Related Work}
\label{section:related-work}

Our work elucidates the connection between three well-studied
techniques.  Hence we can only give a brief overview of the vast
amount of relevant literature.

Since it was observed~\cite{SSS00} that $k$-\emph{induction} for
finite state systems (e.g. hardware circuits) can be done by using an
(incremental) SAT solver \cite{ES03b}, it has become more and more
popular also in the software community as a tool for safety proofs.
Using SMT solvers, it has been applied to Lustre models~\cite{HT08}
(monolithic transition relations) and C programs~\cite{DHKR11}
(multiple and nested loops).

The idea of synthesising abstractions with the help of solvers can be
traced back to predicate abstraction~\cite{GS97}; Reps et
al.~\cite{RSY04} proposed a method for symbolically computing best
abstract transformers;
these techniques were later refined~\cite{BKK11,KGT11,TR12}
for application to various template domains.
Using binary search for optimisation
in this context 
was proposed by Gulwani et al.~\cite{GSV08}.
Similar techniques using LP solving for optimisation 
originate from strategy iteration~\cite{GS07}.
Recently, SMT modulo optimisation~\cite{ST12,LAK+14} techniques
were proposed that foster application to invariant generation
by optimisation.

$k$-induction often requires additional invariants
to succeed, which can be obtained by abstract interpretation.
For example, Garoche et al.~\cite{GKT13} use SMT solving to infer
intermediate invariants over templates for the use in $k$-induction of
Lustre models.  As most of these approaches (except~\cite{BKK11}),
they consider (linear) arithmetic over rational numbers only, whereas
our target are C programs with bit-vectors (representing machine
integers, floating-point numbers, etc).
Moreover, they do not exploit the full power of the approach because
they compute only 1-invariants instead of $k$-invariants.
Another distinguishing feature of our algorithm is that it operates on
a single logical representation and hence enables maximum information
reuse by incremental SAT solving using a single solver.

Formalising program analysis problems such as invariant inference in
second order logic and suggesting to solve these formulae with generic
solvers has been considered by~\cite{GLPR12}.
In this paper we provide an implementation that solves 
the second order formula describing the invariant inference problem
by reduction to quantifier elimination of a first order formula.
Our approach can also solve other problems stated in~\cite{GLPR12},
e.g., termination, by considering different abstract domains, e.g., for
ranking functions.

\section{Conclusions}

This paper presents \algorithmName\ and shows that it can simulate
incremental BMC, $k$-induction and classical, over-approximating
abstract interpretation.  Experiments performed with an
implementation, \systemName, show that it is not only ``more''
complete than each individual technique -- but it also suggests that
it is stronger than their na\"ive combination.  In other words, the
components of the algorithm synergistically interact and enhance each
other.
Moreover, our combination enables a clean, homogeneous, tightly
integrated implementation rather than a loose, heterogeneous
combination of isolated building blocks or a pipeline of techniques
where each only strengthens the next.

There are many possible future directions for this work.  Enhancing
\systemName\ to support additional kinds of templates, possibly
including disjunctive template and improving the optimisation
techniques used for quantifier elimination is one area of interest.
In another direction, \algorithmName\ could be enhance to support
function modular, intraprocedural, thread modular and possibly
multi-threaded analysis.  Automatic refinement of the template domains
is another tantalising possibility.

\bibliographystyle{splncs03}
\bibliography{biblio}

\rronly{
\newpage
\appendix
\section{A Worked Example}\label{sec:running}

We explain the {\algorithmName} algorithm step by step on the
following example:

{\small
\begin{lstlisting}
void main() {
  int w=0,x,y,z;  
  __CPROVER_assume(x==y && y==z && -10<=x && x<0);
  while(1) {
    z = -y;
    y = -x;
    w++;
    x = x + w;
    if(w%2 != 1) w /= 3;
    if(x>=10) x = y = z = 0;
    assert(x<=z+3);
} }
\end{lstlisting}}

\paragraph{SSA construction.} We first perform a lighweight static
analysis in order to translate to program into our SSA form:

{\small
\begin{verbatim}
w#0 == 0
guard#1 == (x#0 < 0 && x#0 == y#0 && y#0 == z#0 && x#0 >= -10)

//loop head
w#phi1 == (guard#ls5 ? w#lb5 : w#0) 
x#phi1 == (guard#ls5 ? x#lb5 : x#0)
y#phi1 == (guard#ls5 ? y#lb5 : y#0)
z#phi1 == (guard#ls5 ? z#lb5 : z#0)

guard#2 == TRUE && guard#1 //in the loop
z#2 == -y#phi1
y#2 == -x#phi1
w#2 == 1 + w#phi1
x#2 == w#2 + x#phi1

guard#3 == (!(w#2 % 2 == 1) && guard#2)
w#3 == w#2 / 3
w#phi4 == (guard#3 ? w#3 : w#2)

guard#4 == ((x#2 >= 10) && guard#2)
z#4 == 0
y#4 == z#4
x#4 == y#4
x#phi5 == (guard#4 ? x#4 : x#2)
y#phi5 == (guard#4 ? y#4 : y#2)
z#phi5 == (guard#4 ? z#4 : z#2)

guard#6 == !TRUE && guard#1 //after loop

guard#2 ==> 3 + z#phi5 >= x#phi5 //assertion
\end{verbatim}
}

It is important to note here that the loop is cut at the end of the
loop body in order to make the SSA acyclic. For this reason, we
replace variables \texttt{w\#phi4}, \texttt{x\#phi5}, \texttt{y\#phi5},
and \texttt{z\#phi5} by free variables \texttt{w\#lb5}, \texttt{x\#lb5},
\texttt{y\#lb5}, and \texttt{z\#lb5} at the loop head.
Since these variables are free we obtain the effect of ``havocking''
these loop variables. The invariants that we compute will constrain
these variables.

\paragraph{Invariant inference}
over the interval domain uses the following guarded template on our example;

{\small
\begin{verbatim}
guard#2 && guard#ls5 ==> 
   w#lb5 <= delta#11 && -((signed __CPROVER_bitvector[33])w#lb5) <= delta#12 && 
   x#lb5 <= delta#21 && -((signed __CPROVER_bitvector[33])x#lb5) <= delta#22 && 
   y#lb5 <= delta#31 && -((signed __CPROVER_bitvector[33])y#lb5) <= delta#32 && 
   z#lb5 <= delta#41 && -((signed __CPROVER_bitvector[33])z#lb5) <= delta#42
\end{verbatim}
}

The casts such as \texttt{(signed \_\_CPROVER\_bitvector[33])w\#lb5)}
are necessary to extend the bitwidth in order to prevent from
arithmetic overflows in template expressions (which would be unsound).
For intervals, the bitwidth extension could be avoided, but our
algorithms are generic for template polyhedra.

Invariant inference using intervals on above program is not very precise. We obtain the following result, which does not allow us to prove the property.
{\small
\begin{verbatim}
guard#2 && guard#ls5 ==> 
   w#lb5 <= 2147483647 && -((signed __CPROVER_bitvector[33])w#lb5) <= 715827882  && 
   x#lb5 <= 9          && -((signed __CPROVER_bitvector[33])x#lb5) <= 2147483648 && 
   y#lb5 <= 2147483647 && -((signed __CPROVER_bitvector[33])y#lb5) <= 2147483648 && 
   z#lb5 <= 2147483647 && -((signed __CPROVER_bitvector[33])z#lb5) <= 2147483648
\end{verbatim}
}

\paragraph{Loop unwinding}
We perform incremental loop unwinding on SSA formula level. 
Since formula construction for incremental loop unwinding is
non-monotonic \cite{SKB+15}, we have to introduce Boolean
variables such as \texttt{enable\#0} which allow us to switch on/off 
certain parts of the formula as needed and use incremental SAT solving
under assumptions \cite{ES03a} to solve these formulae efficiently.

{\small
\begin{verbatim}

w#0 == 0

//loop head of 0th unwinding
enable#0 ==> (guard#1%0 == (x#0 < 0 && x#0 == y#0 && y#0 == z#0 && x#0 >= -10))
enable#0 ==> (w#phi1%0 == (guard#ls5%0 ? w#lb5%0 : w#0))
enable#0 ==> (x#phi1%0 == (guard#ls5%0 ? x#lb5%0 : x#0))
enable#0 ==> (y#phi1%0 == (guard#ls5%0 ? y#lb5%0 : y#0))
enable#0 ==> (z#phi1%0 == (guard#ls5%0 ? z#lb5%0 : z#0))

//last unwinding
guard#2%0 == (TRUE && guard#1%0)
z#2%0 == -y#phi1%0
y#2%0 == -x#phi1%0
w#2%0 == 1 + w#phi1%0
x#2%0 == w#2%0 + x#phi1%0

guard#3%0 == (!(w#2%0 % 2 == 1) && guard#2%0)
w#3%0 == w#2%0 / 3
w#phi4%0 == (guard#3%0 ? w#3%0 : w#2%0)

guard#4%0 == ((x#2%0 >= 10) && guard#2%0)
z#4%0 == 0
y#4%0 == z#4%0
x#4%0 == y#4%0
x#phi5%0 == (guard#4%0 ? x#4%0 : x#2%0)
y#phi5%0 == (guard#4%0 ? y#4%0 : y#2%0)
z#phi5%0 == (guard#4%0 ? z#4%0 : z#2%0)

//merge variables from various loop exits
enable#1 ==> (guard#1 == guard#1%0)
enable#1 ==> (w#phi1 == w#phi1%0)
enable#1 ==> (x#phi1 == x#phi1%0)
enable#1 ==> (y#phi1 == y#phi1%0)
enable#1 ==> (z#phi1 == z#phi1%0)

guard#34 == (!TRUE && guard#1) //after loop exit

guard#2%0 ==> 3 + z#phi5%0 >= x#phi5%0 //assertion
\end{verbatim}
}

For a further iteration we add the following. Note that we unwind
backwards by inserting new unwindings before the old ones.
Also note that formula unwinding generates an exponential blow-up of
the SSA formula in the depth of loop nesting.

{\small
\begin{verbatim}
//loop head of 1st unwinding
enable#1 ==> (guard#1%1 == (x#0 < 0 && x#0 == y#0 && y#0 == z#0 && x#0 >= -10))
enable#1 ==> (w#phi1%1 == (guard#ls5%1 ? w#lb5%1 : w#0))
enable#1 ==> (x#phi1%1 == (guard#ls5%1 ? x#lb5%1 : x#0))
enable#1 ==> (y#phi1%1 == (guard#ls5%1 ? y#lb5%1 : y#0))
enable#1 ==> (z#phi1%1 == (guard#ls5%1 ? z#lb5%1 : z#0))

guard#2%1 == (TRUE && guard#1%1)
z#2%1 == -y#phi1%1
y#2%1 == -x#phi1%1
w#2%1 == 1 + w#phi1%1
x#2%1 == w#2%1 + x#phi1%1

guard#3%1 == (!(w#2%1 % 2 == 1) && guard#2%1)
w#3%1 == w#2%1 / 3
w#phi4%1 == (guard#3%1 ? w#3%1 : w#2%1)

guard#4%1 == ((x#2%1 >= 10) && guard#2%1)
z#4%1 == 0
y#4%1 == z#4%1
x#4%1 == y#4%1
x#phi5%1 == (guard#4%1 ? x#4%1 : x#2%1)
y#phi5%1 == (guard#4%1 ? y#4%1 : y#2%1)
z#phi5%1 == (guard#4%1 ? z#4%1 : z#2%1)

//stitch together 0th and 1st unwinding
enable#1 ==> (guard#1%0 == guard#2%1)
enable#1 ==> (w#phi1%0 == w#phi4%1)
enable#1 ==> (x#phi1%0 == x#phi5%1)
enable#1 ==> (y#phi1%0 == y#phi5%1)
enable#1 ==> (z#phi1%0 == z#phi5%1)

//merge variables from various loop exits
enable#1 ==> (guard#1 == (!guard#2%1 ? guard#1%1 : guard#1%0))
enable#1 ==> (w#phi1 == (!guard#2%1 ? w#phi1%1 : w#phi1%0))
enable#1 ==> (x#phi1 == (!guard#2%1 ? x#phi1%1 : x#phi1%0))
enable#1 ==> (y#phi1 == (!guard#2%1 ? y#phi1%1 : y#phi1%0))
enable#1 ==> (z#phi1 == (!guard#2%1 ? z#phi1%1 : z#phi1%0))

guard#2%1 ==> 3 + z#phi5%1 >= x#phi5%1 //assertion

\end{verbatim}
}

Obviously, unwinding further does not help for this example. This is
why IBMC will not prove the property.

\paragraph{$k$-induction.}
For proving that the property is $k$-inductive, we assume the property
for each unwinding $j<k$ by adding \texttt{guard\#ls5\%}$j$\texttt{ \&\&
  (guard\#2\%}$j$\texttt{ ==> 3 + z\#phi5\%}$j$\texttt{ >=
  x\#phi5\%}$j$\texttt{)} to the formula.  However, the property is
not $k$-inductive on this example.

\paragraph{\algorithmName} additionally uses $k$-inductive invariants.
We infer the following invariant, which, together with the assumptions
from $k$-induction above, allows us to prove the property for $k=2$ on
this example.
{\small
\begin{verbatim}
guard#2%2 && guard#ls5%2 ==> 
   w#lb5%2 <= 1 && -((signed __CPROVER_bitvector[33])w#lb5%2) <= 0  && 
   x#lb5%2 <= 9 && -((signed __CPROVER_bitvector[33])x#lb5%2) <= 10 && 
   y#lb5%2 <= 7 && -((signed __CPROVER_bitvector[33])y#lb5%2) <= 10 && 
   z#lb5%2 <= 6 && -((signed __CPROVER_bitvector[33])z#lb5%2) <= 10
\end{verbatim}
}

\newpage
\section{Further Results}\label{sec:further}

In addition to Table~\ref{tab:results}, Table~\ref{tab:results2} gives
results for an extension of CPAchecker supporting k-induction discussed in
the research report~\cite{BDW15}.  They use a classical abstract interpreter
to generate auxiliary invariants.  Their run times are similar to the
CPAchecker SVCOMP-15 version, but much less complete regarding proofs than
{\algorithmName}.  They only use 1-invariants instead of $k$-invariants, yet
they use increasingly more precise abstract domains.  Moreover, the
extension seems to be still under development as the number of false proofs
and alarms suggests.

\begin{table}
\centering
\begin{tabular}{|l|ccc|c|c|c|c|c|}
\hline
& IBMC & $k$-induction & AI & portfolio & ~\algorithmName\ ~ &
                                                               CPAchecker & ESBMC & CPAchecker-k-ind\\
\hline
counterexamples& \bf 38 & \bf 38 & 17 & \bf 38 & \bf  38 & 36 &  35 & 36 \\
proofs                 & 36 & 49 & 30 & 51 & 52 & 59 &  \bf 91 & 45\\
false proofs        &   0 &   0 &   0 &   0 &  0 &    2   & 12 & 3\\
false alarms        &   2 &   2 &   0 &   2 &   2 &   2   & 0 & 22\\
inconclusive       &   0 &   0 & 93 &   0 &   0 &   4    & 2 & 1\\
timeout               & 65 & 53 &   3 & 50 & 51& 38   &   2 & 36\\
memory out        &   2 &   1 &   0 &  2 &   0 &  2   &   1 & 0\\
total runtime    & 17.1h & 13.8h & 0.89h & 13.3h & 13.2h & 10.9h & 0.54h & 9.1h\\

\hline
\end{tabular}~\\[1ex]
\caption{Comparison between {\algorithmName}, the algorithms it subsumes,
the portfolio, CPAchecker (SVCOMP'15), ESBMC and CPAchecker (k-induction). The rows false alarms and false proofs 
indicate soundness bugs of the tool implementations.
\label{tab:results2}}
\end{table}

}

\end{document}